# Thermo-quantum diffusion

Roumen Tsekov
DWI, RWTH, 52056 Aachen, Germany

A new approach to thermo-quantum diffusion is proposed and a nonlinear quantum Smoluchowski equation is derived, which describes classical diffusion in the field of the Bohm quantum potential. A nonlinear thermo-quantum expression for the diffusion front is obtained, being a quantum generalization of the classical Einstein law. The quantum diffusion at zero temperature is also described and a new dependence of the position dispersion on time is derived. A stochastic Bohm-Langevin equation is also proposed.

The microscopic origin of diffusion is the Brownian motion of the diffusing particles in the surrounding medium. It is named after Robert Brown who has observed first the permanent irregular motion of colloidal particles in water. The theory of Brownian motion has, however, an impact to Science much greater than that of its original object. It is a predecessor of modern kinetic theories in physics (Risken 1996) and of the theory of stochastic differential equations in mathematics (Gardiner 2004). A century ago Langevin (1908) has introduced the first stochastic equation in physics

$$m\ddot{R} + b\dot{R} = X \tag{1}$$

describing the force balance of a Brownian particle with mass $m$. Here $R$ is the Brownian particle coordinate, $b$ is its friction coefficient and $X$ is a random force with zero mean value. Langevin has also anticipated that the Brownian particle coordinate and the Langevin force $X$ are not correlated, which is confirmed later by the advanced theory of stochastic equations. Nowadays, it is well known that the Langevin force is a Gaussian white noise with constant spectral density $S_{XX} = 2bk_B T$ (Gardiner 2004).

Since the Brownian motion is a stochastic process, one needs a statistical approach to relate it to diffusion. The probability density to find the Brownian particle at a given point $r$ at time $t$ can be generally expressed as $P \equiv <\delta(r-R)>$, where the brackets indicate statistical average. In the classical diffusion description $P$ is proportional to the local concentration of the diffusing particles. Differentiating $P$ twice on time one can easily derive the following equation

$$m\partial_t^2 P = \nabla \cdot [\nabla \cdot \Pi - <m\ddot{R}\delta(r-R)>] \tag{2}$$

where the local pressure tensor is introduced via the relation $\Pi \equiv <m\dot{R}\dot{R}\delta(r-R)>$. Expressing now the particle acceleration in Eq. (2) from Eq. (1) yields the following equation

$$m\partial_t^2 P + b\partial_t P = \nabla \cdot (\nabla \cdot \Pi) \qquad (3)$$

Note that the Langevin force does not contribute directly to Eq. (3), since it is not correlated to the Brownian particle position, i.e. $<X\delta(r-R)> = <X>P = 0$, but it contributes indirectly via the pressure tensor. In the high friction limit the first inertial term in Eq. (3) is negligible as compared to the second frictional one. Also the pressure tensor acquires the well-known form $\Pi = <m\dot{R}\dot{R}>P = k_B TPI$, since the kinetic energy of the Brownian particle is not correlated to the particle coordinate and its mean value is proportional to temperature $T$. Introducing this ideal gas expression for $\Pi$ in Eq. (3) leads to the classical diffusion equation

$$\partial_t P = D\nabla^2 P \qquad (4)$$

where $D \equiv k_B T/b$ is the Einstein diffusion constant.

The problem of the thermo-quantum diffusion is how the classical diffusion equation (4) will change if the Brownian particle is a quantum one. At the time when Schrödinger has proposed his famous equation, Madelung (1927) has demonstrated that the Schrödinger equation can be transformed in hydrodynamic form. As a result the quantum effects are included completely into a quantum component $\Pi_Q = -(\hbar^2/4m)P\nabla\nabla \ln P$ of the pressure tensor (Nassar 1985). Therefore, in the case of thermo-quantum diffusion one can present the total pressure tensor in the form of the following superposition (Ancona and Iafrate 1989)

$$\Pi = k_B TPI - (\hbar^2/4m)P\nabla\nabla \ln P \qquad (5)$$

Introducing this expression in Eq. (3) and neglecting again the inertial term yields the following thermo-quantum diffusion equation

$$\partial_t P = \nabla \cdot (P\nabla Q/b + D\nabla P) \qquad (6)$$

where $Q \equiv -\hbar^2 \nabla^2 \sqrt{P}/2m\sqrt{P}$ is the Bohm quantum potential (Bohm 1952, Carroll 2007).

Equation (6) is, in fact, a Smoluchowski equation describing classical diffusion in the field of the quantum potential. Due to $Q$ it is nonlinear. The solution of Eq. (6) is a Gaussian distribution density with dispersion satisfying the following evolution equation (Tsekov 2007)

$$\partial_t \sigma^2 = 2D(1 + \lambda_T^2 / \sigma^2) \tag{7}$$

where $\lambda_T \equiv \hbar / 2\sqrt{mk_B T}$ is the thermal de Broglie wavelength. One can easily integrate Eq. (7) to obtain (Tsekov 2009)

$$\sigma^2 - \lambda_T^2 \ln(1 + \sigma^2 / \lambda_T^2) = 2Dt \tag{8}$$

This expression describes the evolution of the front of the thermo-quantum diffusion. For large times Eq. (8) tends asymptotically to the Einstein law $\sigma^2 = 2Dt$, which holds at any time in the classical limit $\hbar \to 0$. For short times the purely quantum expression $\sigma^2 = \hbar\sqrt{t/mb}$ follows. It holds at any time for $T \to 0$. This expression puts forward also an interesting correspondence to the single-file diffusion (Levitt 1973). Note that the transition time $\lambda_T^2 / D$ between quantum and classical diffusions could be large at low temperatures. The pressure tensor from Eq. (5) acquires the form $\Pi = (k_B T + \hbar^2 / 4m\sigma^2) PI$. Therefore, the momentum dispersion of the quantum Brownian particle $mk_B T + \hbar^2 / 4\sigma^2$ satisfies the Heisenberg uncertainty principle at any time.

The quantum potential is not a potential in the traditional sense, since it depends on the probability density. It drives the quantum diffusion, which is evident from Eq. (6) at zero temperature

$$\partial_t P = \nabla \cdot (P \nabla Q) / b = D_Q \nabla^2 P \tag{9}$$

This equation describes the spreading of a Gaussian wave packet in dissipative environment at $T = 0$. In the derivation of the last expression the Gaussian character of $P$ is employed. As seen, the quantum potential term is equivalent to a diffusion one with a dispersion-dependent diffusion coefficient $D_Q \equiv \hbar^2 / 4mb\sigma^2$. It is appealing to recognize that the thermal diffusion is driven by the Shannon information, while the mean value of the quantum potential is proportional to the Fisher information stored in the probability density (Carroll 2007, Garbaczewski 2008).

It is interesting to find a stochastic equation leading to Eq. (6). A possible candidate is the following Langevin equation

$$m\ddot{R} + b\dot{R} + \nabla U + \nabla Q = X \tag{10}$$

which represents stochastic Bohmian dynamics. For generality, an external potential $U$ is also added to Eq. (10). If this Bohm-Langevin equation is further coupled to Bohmian trajectories via $\dot{R} = V(R,t)$ and $\partial_t P = -\nabla \cdot (PV)$ (Bohm 1952) it leads to the Schrödinger-Langevin equation (Nassar 1985). Our opinion is, however, that the probability density $P$ and the wave function cannot be considered as stochastic quantities. Hence, Eq. (10) is more a mean-field Langevin equation, which is applicable only in the coordinate space due to the Bohm quantum potential. Since the latter accounts explicitly for the quantum effects, the trajectory $R$ must be further described by Eq. (10) as classical one. The spectral density $S_{XX} = 2bk_B T$ of the Langevin force is constant again, since Eq. (10) describes the Brownian motion of a quantum particle in a classical environment. A possibility to generalize the present theory to the case of a quantum Brownian particle in a quantum environment is to employ the spectral density $S_{XX} = b\hbar\omega\coth(\hbar\omega/2k_B T)$ following from the quantum fluctuation-dissipation theorem. It is clear that this expression accounts for quantum effects in the surrounding only, because the included Planck constant appears solely from the bath Hamiltonian. As noted before Eq. (10) is not a standard stochastic equation, since the quantum potential depends on probability density. Hence, it is necessarily bounded to the corresponding Smoluchowski equation

$$\partial_t P = \nabla \cdot [P\nabla(U+Q)/b + D\nabla P] \tag{11}$$

which replaces the continuity equation $\partial_t P = -\nabla \cdot (PV)$ in the de Broglie-Bohm theory. Now the hydrodynamic velocity $V = -\nabla(U + Q + k_B T \ln P)/b$ is averaged over the bath fluctuations and the logarithmic term represents the Boltzmann-Shannon entropy.

The known density gradient model (Ancona and Tiersten 1987, Degond et al 2007) used in semiconductors is based on Eq. (11) with a reduced quantum potential. Let us examine it on a harmonic oscillator with potential $U = m\omega_0^2 r^2/2$, where $\omega_0$ is the oscillator own frequency. The probability density is Gaussian again with dispersion satisfying the following equation

$$\partial_t \sigma^2 = 2D(1 + \lambda_T^2/\sigma^2 - \beta m\omega_0^2 \sigma^2) \tag{12}$$

where $\beta \equiv 1/k_B T$ is the inverse temperature. This equation was recently solved (Messer 2008). Although the corresponding equilibrium dispersion

$$\sigma_e^2 = [1 + \sqrt{1 + (\beta\hbar\omega_0)^2}]/2\beta m\omega_0^2 \tag{13}$$

possesses correct limits at zero and infinite temperatures, it differs from the exact expression

$$\sigma_e^2 = (\hbar/2m\omega_0)\coth(\beta\hbar\omega_0/2) \tag{14}$$

This indicates that Eq. (11) is approximate. The problem appears from the temperature dependence of the quantum potential. Thus, the quantum entropic effect is accounted twice in Eq. (10): first in Q via $P(T)$ and than in the thermal fluctuations of Q via R. Therefore, the exact quantum force is the gradient of the local quantum free energy potential $F_Q \equiv Q - TS_Q$, where the contribution of the quantum thermal entropy $S_Q \equiv -\partial_T F_Q$ is subtracted from the quantum potential. Integrating this Gibbs-Helmholtz relation $\partial_\beta(\beta F_Q) = Q$ yields

$$F_Q = k_B T \int_0^\beta Q d\beta \tag{15}$$

Thus, the density functional stochastic Bohm-Langevin equation acquires the improved form

$$m\ddot{R} + b\dot{R} + \nabla U + \nabla F_Q = X \tag{16}$$

At low temperatures $F_Q$ tends to Q since the entropic effect becomes negligible. Hence, one can consider Eq. (10) as the low temperature limit of Eq. (16). The nonlinear quantum Smoluchowski equation corresponding to Eq. (16) reads

$$\partial_t P = \nabla \cdot [P\nabla(U + F_Q)/b + D\nabla P] = D\nabla \cdot (P\nabla \int_0^\beta \frac{1}{\sqrt{P}} \hat{H}_\beta \sqrt{P} d\beta) \tag{17}$$

which is also derived from a thermodynamically enhanced nonlinear Schrödinger equation (Tsekov 1995, 2008). For compactness, the thermo-quantum Hamiltonian operator

$$\hat{H}_\beta \equiv -(\hbar^2/2m)\nabla^2 + U + \partial_\beta + \partial_\beta^+ \tag{18}$$

is introduced, which is also Hermitian. Since the equilibrium solution of Eq. (17) is the quantum canonical Gibbs distribution (Tsekov 1995, 2008), the equilibrium dispersion of the harmonic oscillator is now the correct one. Hence, Eq. (17) describes properly the thermo-quantum diffusion in the field of an arbitrary potential as well as the equilibrium. It could be especially useful for studies of the thermo-quantum dynamics of the tunneling effect at arbitrary temperature. In the special case of $T = 0$ Eq. (17) reduces to

$$\partial_t P = \nabla \cdot [P\nabla(U + Q)/b] \tag{19}$$

which describes purely quantum diffusion in the field of potential *U*. For a harmonic oscillator the dispersion is given by

$$\sigma^2 = (\hbar/2m\omega_0)\sqrt{1-\exp(-4m\omega_0^2 t/b)} \tag{20}$$

At short time it reduces to the free particle expression $\sigma^2 = \hbar\sqrt{t/mb}$, while at infinite time the equilibrium dispersion is equal to $\sigma_e^2 = \hbar/2m\omega_0$. As seen, the quantum effects do not affect the relaxation time but change the law.

Since the probability density of the harmonic oscillator is Gaussian, the Bohm-Langevin equation (16) can be rewritten in the form

$$m\ddot{R} + b\dot{R} + (m\omega_0^2 - k_B T \int_0^\beta \frac{\hbar^2}{4m\sigma^4} d\beta)R = X \tag{21}$$

As seen, it corresponds to an oscillator with an effective time and temperature dependent spring constant. At large times one can replace the dispersion in the integral above by the equilibrium expression from Eq. (14) to obtain after integration on $\beta$

$$m\ddot{R} + b\dot{R} + (k_B T/\sigma_e^2)R = X \tag{22}$$

This equation describes the equilibrium fluctuations of the oscillator. One can derive easily from Eq. (22) the following expression for the oscillator spectral density

$$S_{RR} = \frac{2bk_B T}{(m\omega^2 - k_B T/\sigma_e^2)^2 + b^2\omega^2} \tag{23}$$

In the high friction limit from Eq. (23) the equilibrium autocorrelation function of the quantum oscillator can be obtained

$$C_{RR} = \sigma_e^2 \exp(-D\tau/\sigma_e^2)I \tag{24}$$

Since the effective spring constant $k_B T/\sigma_e^2$ is smaller than the classical one $m\omega_0^2$, the quantum correlation time $\sigma_e^2/D$ is always larger than the classical $b/m\omega_0^2$. The exponential decay from Eq. (24) is in accordance to the Doob theorem for Gaussian Markov processes (Gardiner 2004). It is interesting that $C_{RR}(T=0) = (\hbar/2m\omega_0)I$ does not depend on time. This means that at zero temperature the equilibrium is static, which is, however, the conclusion of the standard Bohmian mechanics as well (Bohm 1952).

Hereafter an alternative derivation of the nonlinear quantum Smoluchowski equation (17) is proposed. It is well known that the equilibrium density matrix $\rho_e$ satisfies the Bloch (1932) equation

$$\partial_\beta (Z\rho_e) = -\hat{H} Z\rho_e \qquad (25)$$

where $Z$ is the canonical partition function and $\hat{H} = \hat{K} + U$ is the particle Hamiltonian with $\hat{K} \equiv -(\hbar^2/2m)\nabla^2$. Since the Brownian particle is not isolated but exchanging energy with the environment, according to thermodynamics the characteristic function at constant temperature and volume is the Helmholtz free energy $F_e = -k_B T \ln Z$. One can formally express the latter from Eq. (25) as follows

$$F_e = k_B T \int_0^\beta \rho_e^{-1}(\hat{H} + \partial_\beta)\rho_e d\beta \qquad (26)$$

Equation (26) is exact at equilibrium. If the system is not in equilibrium the density matrix $\rho$ will differ from the equilibrium one. However, if the deviation is not very large, one could expect the following extension of Eq. (26) to the non-equilibrium state

$$F(r,t) = k_B T \int_0^\beta [\rho^{-1}(\hat{H} + \partial_\beta)\rho]_{r'=r} d\beta \qquad (27)$$

Now the local free energy $F$ depends on the particle coordinate and time but reduces to $F_e$ at equilibrium. Equation (27) can be further developed in the form

$$F = k_B T \int_0^\beta (\rho^{-1}\hat{K}\rho)_{r'=r} d\beta + U + k_B T \ln(P/P_{\beta=0}) \qquad (28)$$

where $P = \rho_{r'=r}$ is the probability density. As seen the non-equilibrium local free energy $F$ corresponds to the thermodynamic expectations, since in addition to the local kinetic and potential energies the last logarithmic term in Eq. (28) represents the local Boltzmann entropy. Our aim is to build up a mean-field theory, where the local free energy is represented as a functional of the local probability density. For this reason one has to approximate the first kinetic potential in Eq. (28). For slow processes and close to equilibrium one expects to get a good approximation if the density matrix $\rho$ is replaced by the geometric average $\sqrt{P(r',t)P(r,t)}$ of the probability densities. In this case the kinetic potential $(\rho^{-1}\hat{K}\rho)_{r'=r}$ reduces to the Bohm quantum potential $Q$. Thus, the non-equilibrium local free energy from Eq. (28) acquires the following density functional form

$$F = F_Q + U + k_B T \ln(P/P_{\beta=0}) = k_B T \int_0^\beta \frac{1}{\sqrt{P}} (\hat{H} + 2\partial_\beta)\sqrt{P} d\beta \tag{29}$$

The equilibrium solutions of Eq. (29) are $P_n = Z^{-1} \exp(-\beta E_n)\phi_n^2(r)$, where $\phi_n$ and $E_n$ are the real eigenfunctions and eigenvalues of the Brownian particle Hamiltonian, i.e. $\hat{H}\phi_n = E_n \phi_n$. Hence, the equilibrium density matrixes $\rho_n = \sqrt{P_n(r')P_n(r)} = Z^{-1}\exp(-\beta E_n)\phi_n(r')\phi_n(r)$ satisfy Eq. (25). If one is not interested which quantum state the particle occupies than one can sum along the states to obtain $\rho_e = \sum \rho_n$, which also satisfies the linear Bloch equation. The gradient of the free energy is the driving force of diffusion. Substituting $F$ from Eq. (29) in the Gibbs-Duhem equation $\nabla \cdot \Pi = P\nabla F$, introducing the result in Eq. (3) and neglecting of the inertial term results again in Eq. (17).

Ancona, M.G., Tiersten, H.F.: Phys. Rev. B **35**, 7959 (1987)
Ancona, M.G., Iafrate, G.J.: Phys. Rev. B **39**, 9536 (1989)
Bloch, F.: Z. Phys. **74**, 295 (1932)
Bohm, D.: Phys. Rev. **85**, 166 (1952)
Carroll, R.: On the Quantum Potential. Arima Publishing, Suffolk (2007)
Degond, P., Gallego, S., Méhats, F.: J. Comp. Phys. **221**, 226 (2007)
Garbaczewski, P.: Cent. Eur. J. Phys. **6**, 158 (2008)
Gardiner, C.W.: Handbook of Stochastic Methods. Springer, Berlin (2004)
Langevin, P.: Comp. Rend. Acad. Sci. (Paris) **146**, 530 (1908)
Levitt, D.G.: Phys. Rev. A **8**, 3050 (1973)
Madelung, E.: Z. Phys. **40**, 322 (1927)
Messer, J.A.: Giess. Elektr. Bibl. 6328 (2008)
Nassar, A.B.: J. Phys. A: Math. Gen. **18**, L509 (1985)
Risken, H.: The Fokker-Planck Equation. Springer, Berlin (1996)
Tsekov, R.: J. Phys. A: Math. Gen. **28**, L557 (1995)
Tsekov, R.: J. Phys. A: Math. Theor. **40**, 10945 (2007)
Tsekov, R.: Int. J. Theor. Phys. **48**, 85 (2009)